\documentclass[12pt]{article}

\pdfoutput=1
\usepackage[latin9]{inputenc}
\usepackage[top=80pt,bottom=85pt,left=67pt,right=67pt]{geometry}
\usepackage{amssymb,amsmath}
\usepackage{xypic,subfigure}
\usepackage{graphicx,color}
\usepackage{float}
\usepackage{cite}
\usepackage[debug,pageanchor=false]{hyperref}
\definecolor{link}{rgb}{.8,.15,.1}
\hypersetup{colorlinks=true,linkcolor=link,citecolor=link,urlcolor=link,linktocpage}

\makeatletter
\@addtoreset{equation}{section}
\makeatother

\renewcommand{\theequation}{\thesection.\arabic{equation}}

\newcommand{\beq}{\begin{equation}}
\newcommand{\eeq}{\end{equation}}
\newcommand{\bea}{\begin{eqnarray}}
\newcommand{\eea}{\end{eqnarray}}

\newcommand{\eq}{\begin{equation}}
\newcommand{\feq}{\end{equation}}
\newcommand{\eqn}{\begin{eqnarray}}
\newcommand{\feqn}{\end{eqnarray}}

\newcommand{\ma}[1]{\mbox{$\mathcal{#1}$}}

\newcommand{\mrm}[1]{\mbox{$\mathrm{#1}$}}

\begin{document}
\begin{titlepage}

\begin{center}

\vskip .5in 
\noindent

{\Large \bf{AdS$_2$ near-horizons, defects and string dualities}}

\bigskip\medskip

Yolanda Lozano$^{a,b}$\footnote{ylozano@uniovi.es},  Nicol\`o Petri$^c$\footnote{petri@post.bgu.ac.il}, Cristian Risco$^{a,b}$\footnote{cristianrg96@gmail.com}  \\

\bigskip\medskip
{\small 

a: Department of Physics, University of Oviedo,
Avda. Federico Garcia Lorca s/n, 33007 Oviedo}

\medskip
{\small and}

\medskip
{\small 

b: Instituto Universitario de Ciencias y Tecnolog\'ias Espaciales de Asturias (ICTEA),\\
Calle de la Independencia 13, 33004 Oviedo, Spain}

\bigskip\medskip
{\small 

c: Department of Physics, Ben-Gurion University of the Negev, Be'er Sheva 84105, Israel.}

\vskip 2cm 

     	{\bf Abstract }
     	\end{center}
     	\noindent
	
We construct a new family of AdS$_2\times S^3\times S^2$ solutions to Type IIB supergravity arising as near-horizon geometries of D1-F1-D3-D5-NS5-D7 brane intersections preserving 4 supersymmetries. We show that a subclass of these solutions asymptotes locally to the AdS$_6\times S^2\times \Sigma_2$ solution to Type IIB supergravity holographically dual to the five dimensional Sp(N) fixed point theory. This suggests that these solutions can be interpreted as D1-F1-D3 line defects within this CFT.  Switching off the D7-branes, we act  with $\text{SL}(2, \mathbb{R})$ to construct a second family of solutions that can be related to an AdS$_3\times S^3\times S^3$ class of M-theory backgrounds describing surface defects within the six dimensional (1,0) SCFT dual to AdS$_7/\mathbb{Z}_k\times S^4$. Finally, using non-Abelian T-duality we construct new classes of AdS$_2\times S^2\times S^2$ solutions to Type IIA supergravity with 4 supercharges and elaborate on their M-theory origin.	


\noindent

\vfill
\eject

\end{titlepage}

\setcounter{footnote}{0}

\tableofcontents

\setcounter{footnote}{0}
\renewcommand{\theequation}{{\rm\thesection.\arabic{equation}}}

\section{Introduction}

Defects play a prominent role in our current understanding of quantum field theories. Moreover, if the QFT in which they are embedded is conformal, holography provides a very powerful tool for their study \cite{Karch:2001cw,Karch:2000gx,DeWolfe:2001pq}.  In this context they are typically understood as operator insertions that realise a deformation of the ambient CFT. In the so-called probe brane approximation the operator insertion is described by introducing appropriate branes in the dual geometry, that can then be studied using standard supergravity techniques. The probe brane approximation breaks down however when the number of defects is large, due to their backreaction on the original geometry. When this  happens it becomes necessary to know the fully backreacted geometry to properly describe the defects holographically. In this scenario the branes that realise the operator insertion intersect with the brane system where the higher dimensional CFT lives, breaking some of the isometries of its dual AdS vacuum and producing a lower dimensional AdS solution in the near-horizon limit. 
These AdS solutions contain non-trivial warpings between the AdS space and the internal manifold and, in many cases, asymptote locally in a certain limit to the higher dimensional AdS vacuum dual to the ambient CFT \cite{Dibitetto:2017tve,Dibitetto:2017klx,Dibitetto:2018iar,Dibitetto:2018gtk}.

A very useful approach to construct AdS solutions dual to defect CFTs is to search for these solutions in lower dimensional supergravities, and then uplift them to ten or eleven dimensions. The reason for this is that in ten or eleven dimensions the parametrisation of the AdS solution often hides the presence of the higher dimensional AdS vacuum, while in low dimensions one can directly search for solutions in which the defect interpretation is manifest. Following this approach AdS$_2$ and AdS$_3$ backgrounds dual to line and surface defects within 5d and 6d CFTs have been  constructed \cite{Dibitetto:2017tve,Dibitetto:2017klx,Dibitetto:2018iar,Dibitetto:2018gtk,Chen:2019qib,Faedo:2020nol,Faedo:2020lyw,Gutperle:2020rty,Gutperle:2022pgw,Gutperle:2022fma}. Constructions alike directly in ten dimensions have been put forward in
\cite{Lozano:2021fkk,Lozano:2022ouq} and also in \cite{Lozano:2020sae,Lozano:2021rmk,Ramirez:2021tkd}.
Notably, in some cases the operators causing the deformation in the higher dimensional CFT have also been identified. Interesting examples are the AdS$_2$ solutions constructed in \cite{Lozano:2021fkk}  and \cite{Lozano:2020sae,Lozano:2021rmk,Ramirez:2021tkd}, conjectured to be dual to baryon vertices in 4d $\mathcal{N}=4$ SYM and the 5d Sp(N) gauge theory \cite{Seiberg:1996bd}, respectively \cite{Lozano:2021xxs}. 

Besides their applications to the holographic description of defects, low dimensional AdS solutions are  interesting on their own right, as they realise the near-horizon geometries of large classes of extremal black holes, and thereby provide the most promising scenarios where to carry out the microscopical description program. For this reason many efforts have been devoted through the years to scan and classify these spaces. However, due to the high dimensionality of the associated internal manifolds a complete classification of these solutions is still missing. Recently there has been remarkable progress in the classification of AdS$_3$ and AdS$_2$ spaces with 4 supersymmetries \cite{Couzens:2017way,Dibitetto:2018ftj,Macpherson:2018mif,Corbino:2018fwb,Lozano:2019emq,Couzens:2019iog,Couzens:2019mkh,Dibitetto:2019nyz,Legramandi:2019xqd,Lozano:2020bxo,Faedo:2020nol,Lozano:2020txg,Passias:2020ubv,Lozano:2020sae,Legramandi:2020txf,Faedo:2020lyw,Lozano:2021rmk,Ramirez:2021tkd,Couzens:2021tnv,Lozano:2021fkk,Macpherson:2022sbs,Couzens:2022agr,Lozano:2022ouq,Lima:2022hji}. This has come along  with significant advances in our understanding of their 2d and 1d dual CFTs \cite{Couzens:2017way,Lozano:2019jza,Lozano:2019zvg,Lozano:2019ywa,Lozano:2020bxo,Faedo:2020nol,Lozano:2020txg,Lozano:2020sae,Faedo:2020lyw,Couzens:2021veb,Ramirez:2021tkd,Lozano:2021fkk,Lozano:2022ouq}, making these perfect settings where the microscopical description program can be implemented. Of special relevance for our studies in this paper is the interpretation of some of these solutions as holographic duals of defect CFTs \cite{Lozano:2019ywa,Faedo:2020nol,Lozano:2020sae,Faedo:2020lyw,Lozano:2021rmk,Ramirez:2021tkd,Lozano:2021xxs,Lozano:2021fkk,Lozano:2022ouq}. In some cases the explicit knowledge of the quiver gauge theories that describe these CFTs in the UV has allowed to identify the defects with concrete low dimensional vector and matter fields inserted in the quiver gauge theories that describe the higher dimensional CFTs in which they are embedded.  

In this work we present new classes of AdS$_2$ solutions with 4 supersymmetries in Type IIB and Type IIA supergravities, and focus on their defect interpretation. We start with the construction of a general class of Type IIB solutions in section \ref{AdS2TypeIIB}. These backgrounds are obtained as near horizon geometries of 1/8-BPS brane intersections consisting on D1-F1-D3 {\it defect branes} introduced in the D5-NS5-D7 {\it background branes} realising the AdS$_6\times S^2\times \Sigma_2$ solution to Type IIB supergravity constructed in \cite{Cvetic:2000cj,Lozano:2012au} in its near-horizon. 
The defect branes are taken to be completely localised within the worldvolume of the orthogonal background branes, which is the crucial requirement\footnote{At least in the case of AdS$_2$ and AdS$_3$ solutions dual to defects preserving 4 supersymmetries, as shown in \cite{Faedo:2020nol,Faedo:2020lyw,Lozano:2021fkk,Lozano:2022ouq}.} that allows to interpret the solutions as supergravity duals of conformal defects. 
 In order to be able to construct the brane intersection we impose a second requirement, namely, that the D7 and the NS5 branes are smeared over a shared transverse direction. This restricts the possible AdS$_6$ solutions arising in the UV asymptotics to the AdS$_6$ background constructed in \cite{Cvetic:2000cj,Lozano:2012au}, which contains an $S^1$ in the internal space and is
 related by (Abelian) T-duality to the Brandhuber-Oz AdS$_6$ solution to massive IIA supergravity  \cite{Brandhuber:1999np}. This technical restriction is ultimately related to the fact that the brane solutions that underlie more general AdS$_6$ geometries in the classification in  \cite{DHoker:2016ujz,DHoker:2017mds,DHoker:2017zwj}, not containing an $S^1$ in the internal space, are not known.
The presence of the special $S^1$ direction allows to relate our new class of AdS$_2$ solutions in Type IIB to the AdS$_2\times S^3$ solutions in Type IIA supergravity constructed in \cite{Lozano:2020sae}, by means of (Abelian) T-duality. These solutions were interpreted as holographic duals of D0-D4'-F1 baryon vertices in the 5d Sp(N) gauge theory, dual to the Brandhuber-Oz solution \cite{Lozano:2021xxs}. Our solutions in Type IIB find an analogous interpretation, this time as D1-D3-F1 baryon vertices. 

In section \ref{SLrotation} we construct another class of solutions to Type IIB supergravity by acting with SL(2,$\mathbb{R}$) on our previous class of AdS$_2$ backgrounds, restricted to the case without D7-branes. This restriction allows us to perform a local analysis, but at the same time spoils the AdS$_6$ asymptotics. This is related to the fact that AdS$_6$ solutions with a transverse $S^1$ other than the one constructed in  \cite{Cvetic:2000cj,Lozano:2012au}, that contains D7-branes, are not known in Type IIB supergravity. In this case we find that the new AdS$_2$ solutions are related by (Abelian) T-duality to the class of AdS$_3\times S^3\times S^2$ solutions to Type IIA supergravity constructed in \cite{Faedo:2020nol}\footnote{And later extended in \cite{Lozano:2022ouq} to include D8-branes.}, further orbifolded by a $\mathbb{Z}_k$ acting on the AdS$_3$. These solutions asymptote locally to the AdS$_7$ solution to massless Type IIA supergravity constructed in \cite{Cvetic:2000cj,Apruzzi:2013yva}, and allow for a defect interpretation within the 6d (1,0) CFT living in a D6-NS5 brane intersection \cite{Faedo:2020nol}.  In section \ref{dualitywebs} we present the whole web of dualities that relate the two classes of solutions in Type IIB connected by S-duality to the AdS$_2\times S^3$ solutions constructed in \cite{Lozano:2020sae} (restricted to the massless case) and the AdS$_3\times S^3\times S^2$ solutions constructed in \cite{Faedo:2020nol}, and include as well their M-theory realisation. 
In section \ref{NATD2} we construct new AdS$_2$ solutions in Type IIA supergravity with 4 supersymmetries by acting with non-Abelian T-duality on the two previous S-dual backgrounds. Even if after the non-Abelian T-duality transformation we lose a clear interpretation of these solutions as near-horizon geometries of brane intersections, we are still able to relate them to a given M-theory intersection. Finally, section \ref{conclusions} contains our conclusions and open directions. We have collected in an Appendix the details of the uplifts of the solutions in section \ref{NATD2} to M-theory.

\section{The  D1-F1-D3-D5-NS5-D7 brane set-up}\label{AdS2TypeIIB}

In this section we construct a new family of $\mrm{AdS}_2$ solutions to Type IIB supergravity preserving $\ma N=4$ supersymmetries. We obtain these solutions as near-horizon geometries of D1-F1-D3 branes ending on the D5-NS5-D7 brane system where the 5d Sp(N) gauge theory lives. 
Such an intersection reproduces a class of $\mrm{AdS}_2\times S^3\times S^2 \times S^1$ geometries foliated over two intervals in the near horizon. 
We show that a subset of non-compact backgrounds within this class flows asymptotically (locally) to the $\mrm{AdS}_6\times S^2 \times \Sigma_2$ vacuum of Type IIB supergravity constructed in \cite{Cvetic:2000cj,Lozano:2012au}. This AdS$_6$ vacuum geometry was obtained acting with (Abelian) T-duality on the Brandhuber-Oz solution to massive Type IIA supergravity  \cite{Brandhuber:1999np}, and is the only explicit solution within the general classification of AdS$_6\times S^2\times \Sigma_2$ solutions in 
\cite{DHoker:2016ujz, DHoker:2017mds, DHoker:2017zwj} with $\Sigma_2$ an annulus (see \cite{Lozano:2018pcp}). This asymptotic property of our AdS$_2$ solutions allows us to interpret them as holographic duals to line defects within the 5d Sp(N) fixed point theory. In support of the aforementioned interpretation we show that they are related by T-duality to the AdS$_2\times S^3\times \text{CY}_2\times I$ solutions constructed in \cite{Lozano:2020sae} (for $\text{CY}_2=\mathbb{R}^4$), which found themselves an interpretation as line defects within the 5d Sp(N) CFT, as shown in
\cite{Faedo:2020nol,Lozano:2020sae,Lozano:2021xxs}.

We start considering the brane intersection depicted in Table \ref{Table:branesIIBD7}, consisting on D1-F1-D3 branes ending on a D5-NS5-D7 system. 
\begin{table}[http!]
\renewcommand{\arraystretch}{1}
\begin{center}
\scalebox{1}[1]{
\begin{tabular}{c||c|cc c c || c  c| c c c}
 branes & $t$ & $\rho$ & $\varphi^1$ & $\varphi^2$ & $\varphi^3$ & $z$ & $\psi$ & $r$ & $\theta^1$ & $\theta^2$  \\
\hline \hline
$\mrm{D}7$ & $\times$ & $\times$ & $\times$  & $\times$ & $\times$ & $-$ & $-$& $\times$ & $\times$ &$\times$ \\
$\mrm{D}5$ & $\times$ & $\times$ & $\times$ & $\times$ & $\times$ & $-$ & $\times$& $-$ & $-$ & $-$  \\
$\mrm{NS}5$ & $\times$ & $\times$ & $\times$  & $\times$ & $\times$ & $\times$ &$-$& $-$ & $-$ &$-$ \\
$\mrm{D}1$ & $\times$ & $-$ & $-$ & $-$ & $-$ & $ - $ &$\times$& $-$ & $-$ & $-$  \\
$\mrm{F}1$ & $\times$ & $-$ & $-$ & $-$ & $-$ & $\times$ &$-$& $-$ & $-$ & $-$  \\
$\mrm{D}3$ & $\times$ & $-$ & $-$ & $-$ & $-$ & $-$ &$-$& $\times$ & $\times$ & $\times$  \\
\end{tabular}
}
\end{center}
\caption{Brane picture describing the intersection of D5-NS5-D7 branes with D1-F1-D3 branes ending on them. This brane set-up preserves 4 supersymmetries, and is thus 1/8-BPS.} \label{Table:branesIIBD7}
\end{table}
Under certain assumptions this gives rise to the first family of solutions to Type IIB supergravity that we construct in this paper, consisting on AdS$_2\times S^3 \times S^2\times S^1$ fibrations over a 2d Riemann surface.

Our assumptions are as follows. We take the D1-F1-D3 branes completely localised within the worldvolume of the orthogonal D5-NS5-D7 system. This requirement is a crucial property that allows to construct supergravity duals to conformal defects (see  \cite{Faedo:2020lyw,Faedo:2020nol,Lozano:2021fkk,Lozano:2022ouq}), for it allows to decouple the field equations of the {\it defect} branes, D1-F1-D3 in this case, from those of the {\it background} branes, D5-NS5-D7 in our current system. The second important assumption that we make is to take the D7 and NS5 charges smeared over a shared transverse direction. This restricts one to the D5-NS5-D7 brane set-up where the 5d Sp(N) gauge theory lives. With this assumption one recovers asymptotically locally the $\mrm{AdS}_6\times S^2\times\Sigma_2$ solution of Type IIB dual to this SCFT, constructed in 
\cite{Cvetic:2000cj,Lozano:2012au}.

The D1-F1-D3-D5-NS5-D7 system is described by the following 10d metric and dilaton
\begin{equation}
\label{brane_metric_D3D1F1D5NS5D7}
\begin{split}
d s_{10}^2 &= H_{\text{D}7}^{-1/2} H_{\text{D}5}^{-1/2} \left[-H_{\text{D}1}^{-1/2} H_{\text{D}3}^{-1/2}H_{\text{F}1}^{-1} \,dt^2+H_{\text{D}1}^{1/2} H_{\text{D}3}^{1/2}  \bigl(d\rho^2+\rho^2ds^2_{S^3}\bigr) \right] \\
&+H_{\text{D}7}^{1/2} H_{\text{D}5}^{1/2}  H_{\text{D}1}^{1/2} H_{\text{D}3}^{1/2}H_{\text{F}1}^{-1}dz^2+H_{\text{D}7}^{1/2} H_{\text{D}5}^{-1/2} H_{\mathrm{NS}5} H_{\text{D}1}^{-1/2} H_{\text{D}3}^{1/2}d\psi^2\\
&+H_{\text{D}7}^{-1/2} H_{\text{D}5}^{1/2} H_{\mathrm{NS}5} H_{\text{D}1}^{1/2} H_{\text{D}3}^{-1/2} \bigl(dr^2+r^2ds^2_{ S^2}\bigr)\,,\\
e^{\Phi}&= H_{\text{D}7}^{-1} H_{\text{D}5}^{-1/2} H_{\mathrm{NS}5}^{1/2} H_{\text{D}1}^{1/2} H_{\text{F}1}^{-1/2}\,.
\end{split}
\end{equation}
We now ask that the D1-F1-D3 defect branes are completely localised in the $\mathbb{R}^4$ parametrised by $\rho$ and the $S^3$, namely $H_{\text{D}1}$, $H_{\text{F}1}$ and $H_{\text{D}3}$ are just functions of the radial coordinate $\rho$. Further, we impose the smearing of the NS5-D7 branes over the $\psi$ direction, that we assume parametrises a circle, namely $H_{\mathrm{NS}5}=H_{\mathrm{NS}5}(r)$ and $H_{\text{D}7}=H_{\text{D}7}(z)$ \footnote{In the absence of D1-F1-D3 branes and T-dualising along the $\psi$ direction the D4-D8-KK system whose near horizon geometry is the $\mrm{AdS}_6$ vacuum of massive IIA  (orbifolded by $\mathbb{Z}_k$ \cite{Bergman:2012kr}) is reproduced. The KK-monopoles arise from the dualisation of the NS5-branes. In the presence of D1-F1-D3 branes an extra D0-F1-D4$'$ bound state ending on the D4-D8-KK system is obtained. $\mrm{AdS}_2$ solutions associated to these brane intersections were constructed in \cite{Dibitetto:2018gtk,Faedo:2020lyw,Lozano:2020sae}, and interpreted as dual to D0-F1-D4' line defects within the 5d Sp(N) fixed point theory.}. Finally, we take completely localised D5-branes, i.e. $H_{\text{D}5}=H_{\text{D}5}(z,r)$. The fluxes corresponding to this charge distribution acquire the form
\begin{equation}
\begin{split}\label{fluxes_D3D1F1D5NS5D7}
H_{(3)} &= -\partial_\rho H_{\text{F}1}^{-1}dt\wedge d\rho\wedge dz+ \partial_r H_{\mathrm{NS}5}\,r^2\,d\psi \wedge\text{vol}_{S^2}\,,\\
F_{(1)} &= H_{\text{D}1}^{-1} H_{\text{F}1}\,\partial_z H_{\text{D}7} \,d\psi\,,\\
F_{(3)} &= -H_{\text{D}7}\partial_\rho H_{\text{D}1}^{-1}dt\wedge d\rho\wedge d\psi-H_{\text{D}7}\,\partial_r H_{\text{D}5}\,r^2 dz\wedge\text{vol}_{ S^2}+H_{\text{F}1} H_{\text{D}3}^{-1} H_{\mathrm{NS}5}\,r^2\,\partial_zH_{\text{D}5}\, d r \wedge\text{vol}_{ S^2}\,,\\
F_{(5)} &=-H_{\text{D}5} H_{\mathrm{NS}5}\,\partial_\rho H_{\text{D}3}^{-1}\,r^2 \,dt\wedge d\rho\wedge dr \wedge \text{vol}_{ S^2}+ H_{\text{D}7}\partial_\rho H_{\text{D}3}\,\rho^3\,\text{vol}_{ S^3}\wedge dz\wedge d\psi\,.
\end{split}
\end{equation}
Given the metric \eqref{brane_metric_D3D1F1D5NS5D7} and the fluxes \eqref{fluxes_D3D1F1D5NS5D7}, the equations of motion and Bianchi identities of Type IIB supergravity decouple in two groups. One group is associated to the D1-F1-D3 defect branes,
\begin{equation} \label{D3D1F1EOM}
\nabla^2_{\mathbb{R}^4_\rho} H_{\text{D}1}=0 \qquad \text{with}\qquad  H_{\text{D}1}=H_{\text{F}1}=H_{\text{D}3}\,,
\end{equation}
and the other to the D5-NS5-D7 background branes, 
\begin{equation}\label{NS5D5D7EOM}
H_{\text{D}7}\nabla^2_{\mathbb{R}^3_r} H_{\text{D}5} + H_{\mathrm{NS}5} \, \partial_z^2 H_{\text{D}5}=0  \,, \quad  \nabla^2_{\mathbb{R}^3_r} H_{\mathrm{NS}5} = 0  \quad \text{and} \quad \partial_z^2H_{\text{D}7}=0.
\end{equation}
If we now pick the following particular solution to \eqref{D3D1F1EOM},
\begin{equation}
 H_{\text{D}1}=1+\frac{q_{\text{D}1}}{\rho^2}\,,
\end{equation}
 and we take the near-horizon limit $\rho\rightarrow 0$, the following family of backgrounds arises\footnote{In order to have AdS$_2$ with unitary radius we rescaled the coordinates as $t\rightarrow 2^{-1} q_{\text{D}1}^{3/2}t$.},
\begin{equation}
\label{brane_metric_D3D1F1D5NS5D7_nh}
\begin{aligned}
ds_{10}^2 &= 4^{-1} q_{\text{D}1}H_{\text{D}7}^{-1/2}H_{\text{D}5}^{-1/2}\left[ds^2_{\text{AdS}_2} +4 ds^2_{S^3} \right] + H_{\text{D}7}^{1/2}H_{\text{D}5}^{1/2}  dz^2+H_{\text{D}7}^{1/2}H_{\text{D}5}^{-1/2}H_{\mathrm{NS}5}  d\psi^2\\
&+H_{\text{D}7}^{-1/2} H_{\text{D}5}^{1/2}H_{\mathrm{NS}5} \left(dr^2 + r^2 ds^2_{S^2}\right) \,, \\
H_{(3)}&=-2^{-1}q_{\text{D}1}^{1/2}\text{vol}_{\text{AdS}_2} \wedge dz+\partial_r H_{\mathrm{NS}5}\,r^2d\psi \wedge \text{vol}_{S^2} \,,\qquad e^{\Phi}=H_{\text{D}7}^{-1}H_{\text{D}5}^{-1/2}H_{\mathrm{NS}5}^{1/2}\,,\\
F_{(1)}&=\partial_zH_{\text{D}7} \,d\psi\,,\\
  F_{(3)}&=-2^{-1}q_{\text{D}1}^{1/2}H_{\text{D}7} \text{vol}_{\text{AdS}_2} \wedge d\psi-H_{\text{D}7} \partial_r H_{\text{D}5}\,r^2dz\wedge\text{vol}_{ S^2} + H_{\mathrm{NS}5}\,r^2\,\partial_zH_{\text{D}5}\, dr\wedge\text{vol}_{ S^2}\,,\\
 F_{(5)} &=-2^{-1}q_{\text{D}1}^{1/2}H_{\text{D}5} H_{\mathrm{NS}5}\,r^2 \,\text{vol}_{\text{AdS}_2}\wedge dr \wedge \text{vol}_{ S^2}-2q_{\text{D}1} H_{\text{D}7}\,\text{vol}_{ S^3}\wedge dz\wedge d\psi\,.
\end{aligned}
\end{equation}
These backgrounds preserve $\ma N=4$ SUSY. The simplest way to infer this is to note that
they are related to the $\ma N=(0,4)$ $\mrm{AdS}_3\times S^2$ solutions constructed in \cite{Faedo:2020lyw}  through a double analytical continuation\footnote{See the solutions (5.13) of \cite{Faedo:2020lyw}.}. 
We thus obtained a class of $\ma N=4$ $\mrm{AdS}_2\times S^3\times S^2 \times S^1\times I_z\times I_r$ geometries defined by the three functions $H_{\text{D}7}(z)$, $H_{\text{D}5}(z,r)$, $H_{\text{NS}5}(r)$ satisfying equations \eqref{NS5D5D7EOM} and describing the dynamics of a D5-NS5-D7 bound state wrapping an AdS$_2\times S^3$ curved geometry.

\subsection{Line defects within $\mrm{AdS}_6\times S^2\times \Sigma_2$ vacua}\label{defects}

In our previous analysis we derived the supergravity solution describing D1-F1-D3 branes ending on a D5-NS5-D7 system and  showed that in the near-horizon limit the brane solution defines a class of $\ma N=4$ $\mrm{AdS}_2\times S^3\times S^2 \times S^1\times I_z\times I_r$ geometries. These backgrounds are defined by the functions $H_{\text{D}7}(z)$, $H_{\text{D}5}(z,r)$, $H_{\text{NS}5}(r)$ solving the equations of motion of the D5-NS5-D7 bound state, given by equation \eqref{NS5D5D7EOM}.
As we also mentioned at the end of the previous section our solutions can be related via double analytic continuation to the $\ma N=(0,4)$ $\mrm{AdS}_3\times S^2$ solutions constructed in \cite{Faedo:2020lyw}.
These solutions originate from D3-D5-NS5 branes ending on a D5-NS5-D7 system, and under certain assumptions can be interpreted as holographic duals to surface defects within the 5d Sp(N) fixed point theory. One can check that the equations describing the D5-NS5-D7 subsystem of our brane set-up,  given by  \eqref{NS5D5D7EOM}, are exactly the same ones that allowed to find such defect interpretation in \cite{Faedo:2020lyw}. Therefore, we can take the same profiles for $H_{\text{D}7}$, $H_{\text{D}5}$ and $H_{\mathrm{NS}5}$ in order to find AdS$_6$ arising in the asymptotics\footnote{For a detailed derivation see section 5.3 of \cite{Faedo:2020lyw}.}. These profiles are given by \cite{Cvetic:2000cj},
\begin{equation}\label{defectH}
 H_{\mathrm{D}5}=1+\frac{q_{\mathrm{D}5}}{\left(4q_{\mathrm{NS}5}r+\frac49 q_{\mathrm{D}7}z^3 \right)^{5/3}}\,,\qquad H_{\mathrm{NS}5}=\frac{q_{\mathrm{NS}5}}{r}\,, \qquad H_{\mathrm{D}7}=q_{\mathrm{D}7}z\,,
\end{equation}
where the parameters $q_{\mathrm{D}5}$, $q_{\mathrm{D}7}$ and $q_{\mathrm{NS}5}$ are the charges of the D5, NS5 and D7 branes.  As in \cite{Faedo:2020lyw}, the $\mrm{AdS}_6\times S^2\times \Sigma_2$ geometry constructed in  \cite{Cvetic:2000cj,Lozano:2012au} comes out after the change of coordinates,
\begin{equation}\label{defectcoord}
 r= 9^{-1}q_{\mathrm{D}7}\,\mu^{3}\,\cos\alpha^{2}\,\qquad \qquad z=q_{\mathrm{NS}5}^{1/3}\,\mu\,\sin\alpha^{2/3}\,,
\end{equation}
with $\mu>0$ and $\alpha \in [0,\frac{\pi}{2}]$. Indeed,
rewriting the backgrounds \eqref{brane_metric_D3D1F1D5NS5D7_nh}, \eqref{defectH} in this parametrisation and taking the $\mu \rightarrow 0$ limit, one obtains,
\begin{equation}\label{AdS6defect}
ds^2_{10}=s^{-1/3}[\overbrace{4^{-1}q_{\text{D}1}\,q_{\text{NS}5}^{2/3}\mu^2\left(ds^2_{\text{AdS}_2} +4 ds^2_{S^3} \right)+\frac{d\mu^2}{\mu^2}}^{\text{locally}\,\,\, \text{AdS}_6\,\,\, \text{geometry}}+\frac49 d\alpha^2+9 q_{\text{NS}5}^2c^{-2}s^{2/3}d\psi^2+9^{-1}c^2ds^2_{S^2}   ]\,,
\end{equation}
where the 6d external part of the metric describes a locally AdS$_6$ geometry with unit radius.
From this expression it is thus manifest that in the $\mu \rightarrow 0$ limit the $\ma N=4$ solutions take the form of a $\mrm{AdS}_6\times S^2\times \Sigma_2$ vacuum, where the Riemann surface $\Sigma_2$ is an annulus parametrised by the coordinates $(\alpha, \psi)$. Note however that $\mrm{AdS}_6$ arises only locally since extra, subleading, fluxes are also present in the solution that break the $\mrm{AdS}_6$ isometries. Note as well that being the internal space in \eqref{AdS6defect} non-compact along the $\mu$ direction an infinite holographic central charge for the dual superconformal quantum mechanics arises. Indeed, substituting in the general expression for the holographic central charge\footnote{We fixed $G_N^{(10)}=8\pi^6$.} for AdS$_2$ (see \cite{Klebanov:2007ws,Macpherson:2014eza,Bea:2015fja}) we find
\begin{eqnarray}\label{generalF}
 c_{\text{hol}}&=&\frac{3}{8\pi^6}\int_{M_8}d^8y\,\sqrt{g_{8}}\,e^{-2\Phi}\\
& \propto & q_{\text{D}1}^{3/2}\,q_{\text{D}5}^3\, \int d\psi\, d\alpha\, d\mu \cos^3\alpha\,\sin^{1/3}\alpha\,\mu^2\,,\nonumber
\end{eqnarray}
where the integration has been performed along the $M_8$ 8d internal manifold of the AdS$_2$ spacetime. In this expression the divergence along the $\mu$ direction (which plays the role of AdS$_6$ radial coordinate) is manifest. This is exactly the situation one would expect for a 1d CFT dual to a conformal defect embedded in a higher dimensional CFT (see  \cite{Dibitetto:2018gtk, Faedo:2020lyw,Lozano:2022ouq}). 

Finally, it is easy to check that the new AdS$_2\times S^3\times S^2\times S^1\times \Sigma_2$ solutions defined by  \eqref{brane_metric_D3D1F1D5NS5D7_nh} are related by T-duality along the $\psi$ direction to the AdS$_2\times S^3\times \text{CY}_2\times I$ solutions to massive IIA supergravity constructed in \cite{Lozano:2020sae}, for $\text{CY}_2=\mathbb{R}^4$. After the duality the $S^2$ and the $\psi$ direction give rise to a second $S^3$, which together with the $r$ direction build up the $\mathbb{R}^4$. As already mentioned, it was shown in  \cite{Dibitetto:2018gtk, Faedo:2020lyw} that these Type IIA solutions describe D0-F1-D4' branes ending on the D4-D8 system. Further to this, the detailed analysis of the dual field theory performed in  \cite{Lozano:2020sae} allowed to interpret the D0-branes as baryon vertices associated to the D8-branes of the background, and the D4'-branes as baryon vertices associated to the D4 branes\footnote{The field theory is described by quiver-like constructions involving different nodes, and therefore different gauge groups, for both the D4 and the D8 branes. It is worth pointing out that in these constructions the D4 and D8 branes turn from colour branes, where the 5d Sp(N) gauge theory lives, to flavour branes, once the defect branes are introduced. The reader is referred to \cite{Lozano:2020sae,Lozano:2021xxs} for more details on this description.}. Analogously, the D1-F1-D3 defect branes present in our AdS$_2$ solutions find an interpretation 
as D1 and D3 baryon vertices for the D7 and D5 background branes underlying the Type IIB AdS$_6$ solution. The T-duality symmetry that relates these constructions guarantees that the 1d quivers constructed in \cite{Lozano:2020sae}, now built out of D1-D3 colour branes and D7-D5 flavour branes, describe 1d QMs that flow in the IR to the SCQMs dual to our solutions. 

\section{An $\mrm{SL(2,\mathbb R)}$ class of $\ma N=4$ AdS$_2$ near-horizons}\label{SLrotation}

In this section we focus on the subclass of solutions associated to the brane intersection depicted in Table \ref{Table:branesIIBD7} in the absence of D7-branes. Acting with a rotation included in the $\mrm{SL}(2,\mathbb R)$ S-duality group of Type IIB supergravity we obtain a covariant class of solutions depending on the parameter associated to the $\mrm{SL}(2,\mathbb R)$ transformation. As usual, since only $\mrm{SL(2,\mathbb Z)}$ is a symmetry of Type IIB string theory, continuous transformations determine new inequivalent backgrounds in the supergravity limit.

The exclusion of D7 branes is required such that a local analysis of $\mrm{SL}(2,\mathbb R)$ rotations can be performed. Note that this leaves the supersymmetries unaltered. 
Globally one is of course free to take the general brane set-up depicted in Table \ref{Table:branesIIBD7} and perform an S-duality transformation involving the D7-branes. We will refrain however from doing this as we are mainly interested in the local, supergravity description. 

Remarkably, the defect interpretation within AdS$_6$ is lost when the D7-branes are excluded. Still, we will be able to find an interesting defect interpretation within a 6d SCFT once the new solutions have been T-dualised to Type IIA and uplifted to M-theory.

We start by introducing the $\mrm{SL}(2,\mathbb R)$ rotation
\begin{equation}
R=
\left(\begin{array}{cc}
\cos \xi & - \sin \xi \\
\sin\xi &   \cos\xi \\
\end{array}\right)\,.
 \end{equation}
 Acting with it on a ``seed" background described by fluxes, dilaton and metric $F_{(n),s}$, $\Phi_{s}$ and $ds^2_{10,s}$, we have
\begin{equation}
 \begin{split}\label{Srotation_fluxes}
&\left(\begin{array}{c}
\hat{F}_{(3)} \\
H_{(3)}\\
\end{array}\right)=\left(\begin{array}{cc}
\cos \xi & - \sin \xi \\
\sin\xi &   \cos\xi \\
\end{array}\right)\left(\begin{array}{c}
F_{(3),s} \\
H_{(3),s}\\
\end{array}\right)\,,\\
 & \tau=\frac{\cos\xi\,\tau_s-\sin\xi}{\sin\xi\,\tau_s+\cos\xi}\,, \qquad F_{(5)}=F_{(5),s}\,,\\
 \end{split}
\end{equation}
where $\tau=C_{(0)}+ie^{-\Phi}$ stands for the axio-dilaton. Even if the seed solution we are going to consider is characterised by a vanishing axion, this transformation generates a non-trivial profile for $C_{(0)}$. This implies that the 3-form flux associated to the rotated solution is given by $F_{(3)}=\hat F_{(3)}-C_{(0)}H_{(3)}$.
Finally, the metric in the string frame transforms as $ds^2_{10}=|\cos\xi+\sin\xi\,\tau|\,ds^2_{10,s}$. 

Taking as seed solution the brane intersection described by \eqref{brane_metric_D3D1F1D5NS5D7} and \eqref{fluxes_D3D1F1D5NS5D7}, with $H_{D7}=1$, and applying the aforementioned rules we obtain
\begin{equation}
\label{brane_metric_D3D1F1D5NS5D7_rotated}
\begin{split}
d s_{10}^2 &= \Delta^{1/2}\biggl[H_{\text{D}5}^{-1/2} \left(-H_{\text{D}1}^{-1/2} H_{\text{D}3}^{-1/2}H_{\text{F}1}^{-1} \,dt^2+H_{\text{D}1}^{1/2} H_{\text{D}3}^{1/2}  \bigl(d\rho^2+\rho^2ds^2_{S^3}\bigr) \right) \\
&+H_{\text{D}5}^{1/2}  H_{\text{D}1}^{1/2} H_{\text{D}3}^{1/2}H_{\text{F}1}^{-1}dz^2+H_{\text{D}5}^{-1/2} H_{\mathrm{NS}5} H_{\text{D}1}^{-1/2} H_{\text{D}3}^{1/2}d\psi^2\\
&+H_{\text{D}5}^{1/2} H_{\mathrm{NS}5} H_{\text{D}1}^{1/2} H_{\text{D}3}^{-1/2} \bigl(dr^2+r^2ds^2_{ S^2}\bigr)\biggr]\,,\\
\Delta =& \ c^2 + \frac{H_{\text{D}5}}{H_{\mathrm{NS}5}}\frac{H_{\text{F}1}}{H_{\text{D}1}}\,s^2 \,,
\end{split}
\end{equation}
where $s=\sin\xi$ and $c=\cos\xi$. The dilaton and the axion $C_{(0)}$ can be obtained from the axio-dilaton and they have the following form,
\begin{equation}
e^{\Phi}=\Delta\,H_{\text{D}5}^{-1/2}H_{\mathrm{NS}5}^{1/2}H_{\text{D}1}^{1/2}H_{\text{F}1}^{-1/2}\,, \qquad C_{(0)} = \Delta^{-1} \left(\frac{H_{\text{D}5}}{H_{\mathrm{NS}5}}\frac{H_{\text{F}1}}{H_{\text{D}1}} -1 \right)\,sc\,.
\end{equation}
In turn, applying \eqref{Srotation_fluxes} to the fluxes \eqref{fluxes_D3D1F1D5NS5D7} we get
\label{fluxes_D3NS5D5NS5D5rotated}
\begin{equation}
\label{brane_fluxes_D3D1F1D5NS5D7_rotated}
\begin{split}
H_{(3)} &= -c\,\partial_\rho H_{\text{F}1}^{-1}dt\wedge d\rho\wedge dz+c\, \partial_r H_{\mathrm{NS}5}\,r^2\,d\psi \wedge\text{vol}_{S^2}-s\,\partial_\rho H_{\text{D}1}^{-1}dt\wedge d\rho\wedge d\psi\\
&-s\,\partial_r H_{\text{D}5}\,r^2 dz\wedge\text{vol}_{ S^2}+s\,H_{\text{F}1} H_{\text{D}3}^{-1} H_{\mathrm{NS}5}\,r^2\,\partial_zH_{\text{D}5}\, d r \wedge\text{vol}_{ S^2}\,,\\
F_{(3)}&=-c\,\Delta^{-1}\partial_\rho H_{\text{D}1}^{-1}dt\wedge d\rho\wedge d\psi+c\,\Delta^{-1}H_{\text{F}1} H_{\text{D}3}^{-1} H_{\mathrm{NS}5}\,r^2\,\partial_zH_{\text{D}5}\, d r \wedge\text{vol}_{ S^2}\\
&-c\,\Delta^{-1}\partial_r H_{\text{D}5}\,r^2 dz\wedge\text{vol}_{ S^2}+s\,\Delta^{-1}\frac{H_{\text{D}5}}{H_{\mathrm{NS}5}}\frac{H_{\text{F}1}}{H_{\text{D}1}}\,\partial_\rho H_{\text{F}1}^{-1}dt\wedge d\rho\wedge dz\\
&-s\,\Delta^{-1}\frac{H_{\text{D}5}}{H_{\mathrm{NS}5}}\frac{H_{\text{F}1}}{H_{\text{D}1}}\partial_r H_{\mathrm{NS}5}\,r^2\,d\psi \wedge\text{vol}_{S^2}\,,\\
F_{(5)} &=-H_{\text{D}5} H_{\mathrm{NS}5}\,\partial_\rho H_{\text{D}3}^{-1}\,r^2 \,dt\wedge d\rho\wedge dr \wedge \text{vol}_{ S^2}+ \partial_\rho H_{\text{D}3}\,\rho^3\,\text{vol}_{ S^3}\wedge dz\wedge d\psi\,.
\end{split}
\end{equation}
The equations of motion and Bianchi identities are preserved by the $\mrm{SL(2,\mathbb R)}$ rotation, so $H_{\text{D}5}$ and $H_{\mathrm{NS}5}$ must still satisfy equation \eqref{NS5D5D7EOM}, with $H_{\text{D}7}=1$. Note that the absence of D7 branes implies however that $H_{\text{D}3}=H_{\text{F}1}\neq H_{\text{D}1}$ and 
\begin{equation}\label{eomdefect}
 \nabla^2_{\mathbb{R}^4_\rho} H_{\text{D}1}=0 \qquad \text{and} \qquad \nabla^2_{\mathbb{R}^4_\rho} H_{\text{F}1}=0,
\end{equation}
are satisfied instead of \eqref{D3D1F1EOM}.
We can then choose the particular solutions
\begin{equation}
 H_{\text{D}1}=1+\frac{q_{\text{D}1}}{\rho^2}\,,\qquad H_{\text{F}1}=1+\frac{q_{\text{F}1}}{\rho^2},
\end{equation}
and proceed to extract the $\rho\rightarrow 0$ limit. In this way we get a new class of $\ma N=4$ $\mrm{AdS}_2\times S^3\times S^2 \times S^1\times I_z\times I_r$ backgrounds to Type IIB supergravity of the form\footnote{As in the previous section we rescaled the coordinates as $t\rightarrow 2^{-1} q_{\text{D}1}^{1/2}q_{\text{F}1}t$ to have AdS$_2$ with unit radius.}
\begin{equation}
\label{brane_metric_D3D1F1D5NS5D7_nh_rotated}
\begin{aligned}
ds_{10}^2 &=4^{-1} \Delta^{1/2}\biggl[ q_{\text{D}1}^{1/2}q_{\text{F}1}^{1/2}H_{\text{D}5}^{-1/2}\left[ds^2_{\text{AdS}_2} +4 ds^2_{S^3} \right] +q_{\text{D}1}^{1/2}q_{\text{F}1}^{-1/2} H_{\text{D}5}^{1/2}  dz^2\\
&+H_{\text{D}5}^{-1/2}H_{\mathrm{NS}5} \,q_{\text{D}1}^{-1/2}q_{\text{F}1}^{1/2} d\psi^2+H_{\text{D}5}^{1/2}H_{\mathrm{NS}5}q_{\text{D}1}^{1/2}q_{\text{F}1}^{-1/2} \left(dr^2 + r^2 ds^2_{S^2}\right)\biggr] \,, \\
e^{\Phi} &= \Delta  H_{\text{D}5}^{-1/2}H_{\mathrm{NS}5}^{1/2} \,q_{\text{D}1}^{1/2}q_{\text{F}1}^{-1/2} \qquad  \text{with}  \qquad  \Delta = c^2 + \frac{q_{\text{F}1}}{q_{\text{D}1}} \, \frac{H_{\text{D}5}}{H_{\mathrm{NS}5}}\,s^2 \,,\\
H_{(3)} &= -2^{-1}c\,q_{\text{D}1}^{1/2}\text{vol}_{\text{AdS}_2} \wedge dz+c\,\partial_r H_{\mathrm{NS}5}\,r^2d\psi \wedge \text{vol}_{S^2}-2^{-1}s\,q_{\text{D}1}^{-1/2}q_{\text{F}1}  \text{vol}_{\text{AdS}_2} \wedge d\psi\\
&-s\, \partial_r H_{\text{D}5}\,r^2dz\wedge\text{vol}_{ S^2} +s H_{\mathrm{NS}5}\,r^2\,\partial_zH_{\text{D}5}\, dr\wedge\text{vol}_{ S^2}\,,\\
F_{(1)}&=sc\,\Delta^{-2}H_{\mathrm{NS}5}^{-1}\frac{q_{\text{F}1}}{q_{\text{D}1}}\biggl [\partial_zH_{\text{D}5}dz+\left(\partial_r H_{\text{D}5}-H_{\mathrm{NS}5}^{-1}H_{\text{D}5}\partial_r H_{\mathrm{NS}5} \right)dr\biggr ]\,,\\
 F_{(3)}&=-2^{-1}c\,\Delta^{-1}q_{\text{D}1}^{-1/2}q_{\text{F}1} \text{vol}_{\text{AdS}_2} \wedge d\psi-c\,\Delta^{-1}\,\partial_r H_{\text{D}5}\,r^2dz\wedge\text{vol}_{ S^2} +\\
 &+c\,\Delta^{-1} H_{\mathrm{NS}5}\,r^2\,\partial_zH_{\text{D}5}\, dr\wedge\text{vol}_{ S^2}+2^{-1}s\,\Delta^{-1}H_{\text{D}5}H_{\mathrm{NS}5}^{-1}q_{\text{F}1}q_{\text{D}1}^{-1/2}\text{vol}_{\text{AdS}_2} \wedge dz\\
 &-s\,\Delta^{-1}H_{\text{D}5}H_{\mathrm{NS}5}^{-1}q_{\text{F}1}q_{\text{D}1}^{-1}\partial_r H_{\mathrm{NS}5}\,r^2d\psi \wedge \text{vol}_{S^2}\,,\\
 F_{(5)} &=-2^{-1}q_{\text{D}1}^{1/2}H_{\text{D}5} H_{\mathrm{NS}5}\,r^2 \,\text{vol}_{\text{AdS}_2}\wedge dr \wedge \text{vol}_{ S^2}-2q_{\text{F}1} \,\text{vol}_{ S^3}\wedge dz\wedge d\psi\,.
\end{aligned}
\end{equation}
Here $H_{\text{D}5}$ and $H_{\mathrm{NS}5}$ must satisfy the equations
\begin{equation}\label{eq-D5NS5}
\nabla^2_{\mathbb{R}^3_r} H_{\text{D}5} + H_{\mathrm{NS}5} \, \partial_z^2 H_{\text{D}5}=0  \qquad  \text{and}  \qquad  \nabla^2_{\mathbb{R}^3_r} H_{\mathrm{NS}5} = 0  \,.
\end{equation}
The above class of solutions describes $(p^\prime,q^\prime)$ strings and D3 branes ending on orthogonal $(p,q)$ 5-branes, and is in this sense more general than the class of solutions constructed in Section \ref{AdS2TypeIIB}. This is reflected by the fact that the D5 and NS5 charges are now distributed along the $(z,\psi,\rho)$ directions while the D1 and F1 charges are mixed along $(z,\psi)$. The interpretation of these solutions should be as holographic duals to D3 baryon vertices introduced in the 5d field theory living in D5-NS5 branes, with F1 (D1) strings in the completely antisymmetric representation of the D5 (NS5) gauge groups stretched between the D3 and the D5 (NS5) branes. It would be interesting to provide a concrete realisation of this set-up, along the lines of \cite{Lozano:2020sae,Lozano:2021rmk,Ramirez:2021tkd,Lozano:2021fkk}.

It will be useful for our constructions in Section \ref{NATD2} to have the explicit form of the solutions \eqref{brane_metric_D3D1F1D5NS5D7_nh_rotated} particularised to $\xi=0,\frac{\pi}{2}$, that is, the two families of solutions in this class that are S-dual to one another. For $\xi=0$ we have
\begin{equation}
	\label{brane_metric_D3D1F1D5NS5D7_nh_0}
	\begin{aligned}
		ds_{10}^2 &=4^{-1} q_{\text{D}1}^{1/2}q_{\text{F}1}^{1/2}H_{\text{D}5}^{-1/2}(ds^2_{\text{AdS}_2} +4 ds^2_{S^3} ) +q_{\text{D}1}^{1/2}q_{\text{F}1}^{-1/2} H_{\text{D}5}^{1/2}  dz^2\\
		&+H_{\text{D}5}^{-1/2}H_{\mathrm{NS}5} \,q_{\text{D}1}^{-1/2}q_{\text{F}1}^{1/2} d\psi^2+H_{\text{D}5}^{1/2}H_{\mathrm{NS}5}q_{\text{D}1}^{1/2}q_{\text{F}1}^{-1/2} \left(dr^2 + r^2 ds^2_{S^2}\right) \,, \\
		e^{\Phi} &= H_{\text{D}5}^{-1/2}H_{\mathrm{NS}5}^{1/2} \,q_{\text{D}1}^{1/2}q_{\text{F}1}^{-1/2} \,,\\
		H_{(3)} &= -2^{-1}\,q_{\text{D}1}^{1/2}\text{vol}_{\text{AdS}_2} \wedge dz+\,\partial_r H_{\mathrm{NS}5}\,r^2d\psi \wedge \text{vol}_{S^2}\,,\\
		F_{(3)}&=-2^{-1} q_{\text{D}1}^{-1/2}q_{\text{F}1} \text{vol}_{\text{AdS}_2} \wedge d\psi-\,\partial_r H_{\text{D}5}\,r^2dz\wedge\text{vol}_{ S^2} + H_{\mathrm{NS}5}\partial_zH_{\text{D}5}\, r^2\, dr\wedge\text{vol}_{ S^2}\,,\\
		F_{(5)} &=-2^{-1}q_{\text{D}1}^{1/2}H_{\text{D}5} H_{\mathrm{NS}5}\,r^2 \,\text{vol}_{\text{AdS}_2}\wedge dr \wedge \text{vol}_{ S^2}-2q_{\text{F}1} \,\text{vol}_{ S^3}\wedge dz\wedge d\psi\,.
	\end{aligned}
\end{equation}
Note that this class of solutions is a generalisation of  the backgrounds \eqref{brane_metric_D3D1F1D5NS5D7_nh} (with $H_{\text{D}7}=1$) where $H_{\text{D}1}\neq H_{\text{F}1}$ and there are therefore both $q_{D1}$ and $q_{F1}$ quantised charges.
In turn, for $\xi=\frac{\pi}{2}$ we have
\begin{equation}
	\label{brane_metric_D3D1F1D5NS5D7_nh_pi_2}
	\begin{aligned}
		d s_{10}^2 =&4^{-1}q_{\text{F}1} H_{\mathrm{NS}5}^{-1/2} (ds_{\text{AdS}_2}^2 + 4 ds_{S^3}^2) + H_{\mathrm{NS}5}^{-1/2} H_{\text{D}5} dz^2+\\
		& + q_{\text{F}1}q_{\text{D}1}^{-1} H_{\mathrm{NS}5}^{1/2} d\psi^2
		+  H_{\mathrm{NS}5}^{1/2}  H_{\text{D}5}(dr^2 +  r^2 ds_{S^2}^2 ) \,,\\
		e^{\Phi} =&  H_{\mathrm{NS}5}^{-1/2}  H_{\text{D}5}^{1/2} q_{\text{F}1}^{1/2} q_{\text{D}1}^{-1/2}\,, \\
		H_{(3)} =& - 2^{-1} q_{\text{F}1}q_{\text{D}1}^{-1/2} \ \text{vol}_{\text{AdS}_2}\wedge d\psi  - \partial_r H_{\mathrm{D}5} r^2 dz \wedge \text{vol}_{S^2}+ H_{\text{NS}5} \partial_z H_{\mathrm{D}5} r^2 \ dr \wedge \text{vol}_{S^2}\,,
		\\
		F_{(3)} =& 2^{-1} q_{\text{D}1}^{1/2} \ \text{vol}_{\text{AdS}_2} \wedge dz -  \partial_r H_{\text{D}5} r^2 \ d\psi \wedge \text{vol}_{S^2}\,,
		\\
		F_{(5)} =& - 2^{-1} q_{\text{D}1}^{1/2}  H_{\text{D}5} H_{\mathrm{NS}5} r^2 \ \text{vol}_{\text{AdS}_2}   \wedge dr  \wedge \text{vol}_{S^2} - 2 q_{\text{F}1}  \text{vol}_{S^3}\wedge dz  \wedge d\psi\,.
	\end{aligned}
\end{equation}

Finally, we can provide a unified expression for the central charge of the whole family of $\mrm{SL}(2,\mathbb R)$ solutions, since this quantity is $\mrm{SL}(2,\mathbb R)$ invariant. Substituting the metric and dilaton of the backgrounds \eqref{brane_metric_D3D1F1D5NS5D7_nh_rotated} in 
\eqref{generalF} we indeed find
\begin{equation}
c_{\text{hol}}=\frac{3}{8\pi^6}\int_{M_8}d^8y\,\sqrt{g_{8}}\,e^{-2\Phi}=\frac{3}{8\pi^6}\, q_{\text{D}1}^{1/2}\,q_{\text{F}1}\,\text{Vol}_{S^3}\text{Vol}_{S^2}\int\,d\psi\,dr\, dz\,r^2 H_{\text{D}5}\, H_{\mathrm{NS}5}\,,
\end{equation}
where the $\xi$-parameter is not present.

\section{Web of dualities and M-theory origin}\label{dualitywebs}

In this section we discuss the Type IIA realisation and M-theory origin of the $\xi=0$ and $\xi=\frac{\pi}{2}$ solutions to Type IIB constructed in the previous section. As we already mentioned the defect interpretation within AdS$_6\times S^2\times \Sigma_2$ is lost. Instead, the solutions allow for an interesting realisation as line defects within the 6d (1,0) CFT dual to 
AdS$_7/\mathbb{Z}_k \times S^4$ in M-theory.

The two S-dual solutions with $\xi=0, \frac{\pi}{2}$  in \eqref{brane_metric_D3D1F1D5NS5D7_nh_0} and \eqref{brane_metric_D3D1F1D5NS5D7_nh_pi_2} are related by T-duality to the AdS$_2\times S^3\times \mathbb{R}^4/\mathbb{Z}_k$ solutions constructed in  \cite{Lozano:2020sae}\footnote{Restricted to the massless case, since we are not allowing for D7-branes.}, and to the AdS$_3/\mathbb{Z}_{k'}\times S^3\times S^2$ solutions to massless IIA supergravity constructed in \cite{Faedo:2020nol}\footnote{With the AdS$_3$ modded out by $\mathbb{Z}_{k'}$, but this is a trivial extension of the solutions in  \cite{Faedo:2020nol}.}, respectively. As shown in  \cite{Faedo:2020nol} these solutions
share a common origin in M-theory, in the form of AdS$_3\times S^3\times S^3/\mathbb{Z}_k$ backgrounds (AdS$_3/\mathbb{Z}_{k'}$ in our case), also classified in  \cite{Faedo:2020nol}. 
These solutions were shown to asymptote to AdS$_7/\mathbb{Z}_k\times S^4$ in the UV.  Our solutions are thus interpreted as duals to line defects in the 6d (1,0) CFT dual to this background, once uplifted to M-theory. 

\begin{figure}[http!]
	\begin{center}
		\scalebox{0.8}[0.8]{ \xymatrix@C-0pc { 
				\text{  } & *+[F-,]{\begin{array}{c} \textrm{M0 - M2 - M5   on M5'  - KK}\vspace{2mm} \\ \textrm{AdS}_3/\mathbb{Z}_{k'}\times S^3 \times \mathbb{R}^4/\mathbb{Z}_k\times I_z \end{array}} \ar[dl]_{\chi} \ar[dr]^{\psi} & \text{  }    \\
		*+[F-,]{\begin{array}{c} \textrm{ D0 - F1 - D4'  on D4-KK } \vspace{2mm} \\ \textrm{AdS}_2\times S^3  \times \mathbb{R}^4/\mathbb{Z}_k \times I_z  \end{array}}\ar@{<->}[d]^{\mathrm{T}_\psi} &  \text{  }  & 
		*+[F-,]{\begin{array}{c} \textrm{wave - D2 - D4  on NS5 - D6} \vspace{2mm} \\ \textrm{AdS}_3/\mathbb{Z}_{k'} \times S^3  \times  S^2\times I_z \times I_r \end{array}}\ar@{<->}[d]^{\mathrm{T}_\chi}\\
		*+[F-,]{\begin{array}{c} \textrm{D1 - F1 - D3  on D5 - NS5 } \vspace{2mm} \\  \textrm{AdS}_2\times S^3  \times  S^2\times S^1_\psi \times I_z \times I_r  \end{array}} &  \ar[l] \text{S} \ar[r] & 
		*+[F-,]{\begin{array}{c} \textrm{F1 - D1 - D3  on NS5 - D5 } \vspace{2mm} \\ \textrm{AdS}_2\times S^3  \times  S^2\times S^1_\chi \times I_z \times I_r  \end{array}}
		}}
	\end{center}
	\caption{Web of dualities that relate the new AdS$_2$ solutions in Type IIB written in \eqref{brane_metric_D3D1F1D5NS5D7_nh_0} and \eqref{brane_metric_D3D1F1D5NS5D7_nh_pi_2} to the Type IIA and M-theory solutions constructed in  \cite{Lozano:2020sae} and \cite{Faedo:2020nol}.}\label{fig:dualities} 
\end{figure}
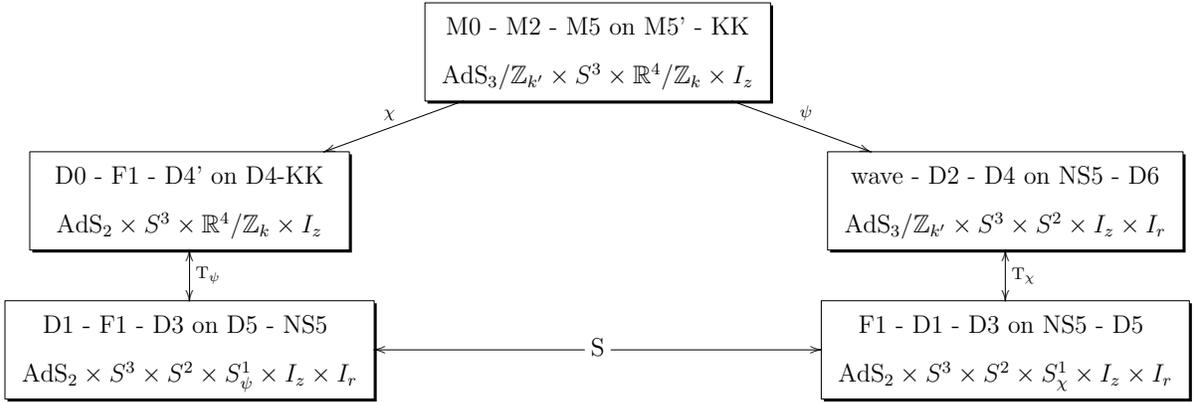

The web of dualities connecting these classes of solutions is depicted in Figure \ref{fig:dualities}, that we now explain in detail.
Starting with the bottom left solution of Type IIB and performing a T-duality along the $S^1_\psi$ circle, an $S^3$ is built up with the $S^1_\psi$ and the $S^2$. This $S^3$ gives rise to an $\mathbb{R}^4/\mathbb{Z}_k$ space together with the $r$-direction. Here the integer $k$ is the number of NS5-branes present in the Type IIB solution, that  become KK-monopoles in Type IIA. This solution in IIA is contained in the class found in \cite{Lozano:2020sae}, for $\text{CY}_2=\mathbb{R}^4/\mathbb{Z}_k$. The corresponding Type IIA brane set-up is described by a D4-KK-F1-D4'-D0 intersection studied in \cite{Dibitetto:2018gtk} and it is depicted in the left-hand side of Table \ref{Table:branesinmasslessIIA2}.
\begin{table}[http!]
\renewcommand{\arraystretch}{1}
\begin{center}
\scalebox{1}[1]{
\begin{tabular}{c||c c|c c c  c | c  c c c  c }
branes & $t$ & $\chi$ &  $\rho$ & $\varphi^1$ & $\varphi^{2}$ & $\varphi^{3}$ & $z$&$\psi$ & $r$ & $\theta^1$ & $\theta^2$   \\
\hline \hline
$\mrm{M}5$' & $\times$ & $\times$ & $\times$ & $\times$ & $\times$ & $\times$ & $-$ & $-$ & $-$ & $-$ & $-$ \\
$\mrm{KK}$ & $\times$ & $\times$ & $\times$ & $\times$ & $\times$ & $\times$ & $\times$ & $\text{ISO}$ & $-$ & $-$  & $-$ \\
$\mrm{M}2$& $\times$ & $\times$ & $-$ & $-$ & $-$ & $-$ & $\times$ & $-$ & $-$ & $-$ & $-$ \\
$\mrm{M}5$ & $\times$ & $\times$ & $-$ & $-$ & $-$ & $-$ & $-$ & $\times$ & $\times$ & $\times$ & $\times$ \\
$\mrm{M}0$ & $\times$ & $\text{ISO}$ &$-$ & $-$ & $-$ & $-$ & $-$ & $-$ & $-$ & $-$  & $-$\\
\end{tabular}
}
\end{center}
\label{Table:branesMtheory}
\end{table}
\begin{table}[h!]
\renewcommand{\arraystretch}{1}
\scalebox{0.82}[0.8]{
\begin{tabular}{c||c|cc c c | c  c c c c}
 branes & $t$ & $\rho$ & $\varphi^1$ & $\varphi^{2}$ & $\varphi^{3}$ & $z$&$\psi$ & $r$ & $\theta^1$ & $\theta^2$\\
\hline \hline
$\mrm{D}4$ & $\times$ & $\times$ & $\times$ & $\times$ & $\times$ & $-$ & $-$ & $-$ & $-$&$-$  \\
$\mrm{KK}$ & $\times$ & $\times$ & $\times$  & $\times$ & $\times$ & $\times$ &$\text{ISO}$& $-$ & $-$ &$-$ \\
$\mrm{F}1$ & $\times$ & $-$ & $-$ & $-$ & $-$ & $\times$ &$-$& $-$ & $-$ & $-$  \\
$\mrm{D}4'$ & $\times$ & $-$ & $-$ & $-$ & $-$ & $-$ &$\times$& $\times$ & $\times$ & $\times$  \\
$\mrm{D}0$ & $\times$ & $-$ & $-$ & $-$ & $-$ & $-$ & $-$ & $-$ & $-$ &$-$ \\
\end{tabular}
\quad
\begin{tabular}{c||cc|c c c  c  |c c c c}
branes &  $t$ &$\chi$& $\rho$ & $\varphi^1$ & $\varphi^{2}$ & $\varphi^{3}$ & $z$  & $r$ & $\theta^1$ & $\theta^2$ \\
\hline \hline
$\mrm{NS}5$ & $\times$ &$\times$& $\times$ & $\times$ & $\times$ & $\times$ & $-$ & $-$ & $-$ & $-$  \\
$\mrm{D}6$ & $\times$ &$\times$& $\times$ & $\times$  & $\times$ & $\times$ & $\times$ & $-$ & $-$ &$-$ \\
$\mrm{D}2$ & $\times$ &$\times$& $-$ & $-$ & $-$ & $-$ & $\times$ & $-$ & $-$ & $-$  \\
$\mrm{D}4$ & $\times$ & $\times$& $-$ & $-$ & $-$ & $-$ & $-$ & $\times$ & $\times$ & $\times$  \\
$\text{W}$ & $\times$ & $$\text{ISO}$$ & $-$ & $-$ & $-$ & $-$ & $-$ & $-$ & $-$&$-$  \\
\end{tabular}
}
\caption{1/8-BPS brane set-ups in M-theory and Type IIA associated to Figure \ref{fig:dualities}. In M-theory one has the intersection of M2-M5-M0 branes ending on M5'-branes with KK monopoles. The reduction to Type IIA can be performed over the coordinates $\chi$ and $\psi$ that parametrise the isometric directions and are respectively associated to the momentum waves M0 and the KK monopoles.} \label{Table:branesinmasslessIIA2}
\end{table}

We already referred  to this T-duality transformation in Section \ref{AdS2TypeIIB}, for the more general situation in which D7-branes were also present. In turn, the uplift of the Type IIA solution to M-theory produces an AdS$_3/\mathbb{Z}_{k'}$ space, built up with the AdS$_2$ and the M-theory circle (parametrised by the $\chi$ coordinate), where $k'$ is the number of F1-strings in the Type IIA solution, that become waves, or units of momentum, in M-theory. The M-theory intersection undelying these solutions is depicted on top of Table \ref{Table:branesinmasslessIIA2} and it is defined by an intersection of M5'-KK-M2-M5-M0 branes. The corresponding class of AdS$_3/\mathbb{Z}_{k'}\times S^3\times \mathbb{R}^4/\mathbb{Z}_k\times I_z$ solutions to M-theory was found in  \cite{Lozano:2020bxo}. In turn, it belongs to the more general class of AdS$_3\times S^3\times S^3/\mathbb{Z}_k\times \Sigma_2$ solutions constructed in \cite{Faedo:2020nol}, in our case orbifolded by $\mathbb{Z}_{k'}$. 
Taking now these solutions as our starting point, but reducing instead along the $S^1_\psi/\mathbb{Z}_k$ Hopf fibre of the $S^3/\mathbb{Z}_k$ contained in $\mathbb{R}^4/\mathbb{Z}_k$, we obtain a solution in Type IIA in the class constructed in \cite{Faedo:2020nol}\footnote{And later generalised to the massive case in \cite{Lozano:2022ouq}.}, with extra $k'$ waves, or units of momentum. The corresponding brane set-up is presented in the right-hand side of Table \ref{Table:branesinmasslessIIA2} and it is given by an intersection of D6-NS5-D4-D2 branes with momentum waves W. T-dualising along the Hopf fibre of the AdS$_3/\mathbb{Z}_{k'}$ subspace we finally arrive at the Type IIB solution shown at the bottom right of the figure, containing $k'$ F1-strings. As expected due to their common M-theory origin, both solutions in Type IIB are related to each other by S-duality.

\section{Non-Abelian T-duals and Type IIA picture} \label{NATD2}

In this section we present new AdS$_2$ solutions to Type IIA supergravity preserving 4 supercharges obtained by performing a non-Abelian T-duality transformation  along the $S^3$ on the two S-dual backgrounds with $\xi=0$ and $\xi=\pi/2$ given in \eqref{brane_metric_D3D1F1D5NS5D7_nh_0} and \eqref{brane_metric_D3D1F1D5NS5D7_nh_pi_2}. These Type IIA backgrounds depend on two defining functions  $H_{\text{D}5}=H_{\text{D}5}(z,r)$, $H_{\mathrm{NS}5}=H_{\mathrm{NS}5}(r)$ satisfying the master equations\footnote{\label{noteq3}In this section we restore the integration constant $q_{\text{D}3}$ associated to D3 defect branes in the S-dual Type IIB backgrounds \eqref{brane_metric_D3D1F1D5NS5D7_nh_0} and \eqref{brane_metric_D3D1F1D5NS5D7_nh_pi_2}. We recall that this parameter was fixed as $q_{\text{D}3}=q_{\text{F}1}$ at the level of the brane solution $H_{\text{D}3}=1+\frac{q_{\text{D}3}}{\rho^2}$, $H_{\text{F}1}=1+\frac{q_{\text{F}1}}{\rho^2}$ in \eqref{brane_metric_D3D1F1D5NS5D7_nh_rotated} by the conditions $\nabla^2_{\mathbb{R}^4_\rho}\,H_{\text{D}3}=\nabla^2_{\mathbb{R}^4_\rho}\,H_{\text{F}1}=0$ and $H_{\text{D}3}=H_{\text{F}1}$ coming from the equations of motion for the defect branes (written in \eqref{eomdefect}). The freedom to keep $q_{\text{D}3}$ unconstrained at the near-horizon is provided by the fact that the condition $H_{\text{D}3}=H_{\text{F}1}$ is a particular realisation of the slightly more general condition $H_{\text{D}3}H'_{\text{F}1}=H'_{\text{D}3}H_{\text{F}1}$, implied by the equations of motion. Outside of the near-horizon these two conditions are equivalent and imply that $q_{\text{D}3}=q_{\text{F}1}$, but in the $\rho \rightarrow 0$ limit the absence of the ``1"  factor in the harmonic functions $H_{\text{D}3}=\frac{q_{\text{D}3}}{\rho^2}$, $H_{\text{F}1}=\frac{q_{\text{F}1}}{\rho^2}$ allows one to avoid any constraint on $q_{\text{D}3}$ in terms of the other integration constants. The AdS$_2$ factor with unitary radius in the metrics of the S-dual solutions in Type IIB is realised by the rescaling of the time direction $t\rightarrow 2^{-1}q_{\text{D}1}^{1/2}q_{\text{D}3}^{1/2}q_{\text{F}1}^{1/2}\,t$.} 
\begin{equation}\label{eq-D5NS5}
\nabla^2_{\mathbb{R}^3_r} H_{\text{D}5} + \frac{q_{\text{F}1}}{q_{\text{D}3}} H_{\mathrm{NS}5} \, \partial_z^2 H_{\text{D}5}=0  \qquad  \text{and}  \qquad  \nabla^2_{\mathbb{R}^3_r} H_{\mathrm{NS}5} = 0  \,.
\end{equation}
Under non-Abelian T-duality the $S^{3}$ of the original background is transformed into an open subset of $\mathbb{R}^3$, parametrised by the radial coordinate $R$ and the 2-sphere ${\tilde S}^2$. 
For $\xi=0$ the new class of non-Abelian T-dual solutions is given by,
\begin{equation}
\label{brane_metric_NATD_0_nh}
\begin{aligned}
ds_{10}^2 &= 4^{-1} q_{\text{D}1}^{1/2}q_{\text{D}3}^{1/2}H_{\text{D}5}^{-1/2}ds^2_{\text{AdS}_2} +q_{\text{D}1}^{1/2}q_{\text{D}3}^{1/2} q_{\mathrm{F}1}^{-1} H_{\text{D}5}^{1/2}  dz^2 +q_{\text{D}1}^{-1/2}q_{\text{D}3}^{1/2}H_{\text{D}5}^{-1/2}H_{\mathrm{NS}5} \, d\psi^2+\\
&+q_{\text{D}1}^{1/2}q_{\text{D}3}^{-1/2}H_{\text{D}5}^{1/2}H_{\mathrm{NS}5} \left(dr^2 + r^2 ds^2_{S^2}\right)+q_{\text{D}1}^{-1/2}q_{\text{D}3}^{-1/2}H_{\text{D}5}^{1/2}4 (dR^2 + H R^2 ds^2_{\tilde{S}^2}) \,, \\
e^{\Phi} &=  8q_{\text{D}1}^{-1/4}q_{\mathrm{F}1}^{-1/2}q_{\text{D}3}^{-3/4}H_{\text{D}5}^{1/4} H_{\mathrm{NS}5}^{1/2} \, H^{1/2}\,,\\
H_{(3)} &= -2^{-1}\,q_{\text{D}1}^{1/2}q_{\text{D}3}^{1/2}q_{\mathrm{F}1}^{-1/2}\text{vol}_{\text{AdS}_2} \wedge dz+\,\partial_r H_{\mathrm{NS}5}\,r^2d\psi \wedge \text{vol}_{S^2}\\
&+\partial_zHR dz\wedge\text{vol}_{\tilde{S}^2}+\partial_rHRdr\wedge \text{vol}_{\tilde{S}^2}+\partial_R((H-1)R)\,\,dR\wedge \text{vol}_{\tilde{S}^2}\,,\\
F_{(2)} &= -4^{-1}q_{\text{D}3} \,dz\wedge d\psi\,,\\
F_{(4)} &= d[2^{-1}q_{\text{D}1}^{-1/2}q_{\mathrm{F}1}^{1/2}q_{\text{D}3}^{1/2}(3/2+(H-1)^{-1})R^2 \text{vol}_{\text{AdS}_2} \wedge d\psi]\\
&+4 q_{\text{D}1}^{-1} HR^3H_{\text{D}5} \, dz \wedge d\psi \wedge \text{vol}_{\tilde S^2}-4^{-1}q_{\text{D}1}H_{\text{D}5} H_{\mathrm{NS}5}\,r^2 \,dz\wedge dr \wedge \text{vol}_{ S^2}+\\
&+r^2R(q_{\mathrm{F}1}q_{\text{D}3}^{-1}H_{\mathrm{NS}5}\,\partial_zH_{\text{D}5}dr-\partial_r H_{\text{D}5}\,dz)\,  \wedge\text{vol}_{ S^2}\wedge dR=\,,
\end{aligned}
\end{equation}
where we have defined,
\begin{equation}
H = \frac{q_{\text{D}1}q_{\text{D}3}}{q_{\text{D}1}q_{\text{D}3} + 16 R^2 H_{\text{D}5}}\,.
\end{equation}
In turn, for $\xi=\frac{\pi}{2}$ we find the new class,
\begin{equation}
\label{brane_metric_NATD_pi/2_nh}
\begin{aligned}
ds_{10}^2 &= 4^{-1} q_{\mathrm{F}1}^{1/2}q_{\text{D}3}^{1/2} H_{\mathrm{NS}5}^{-1/2}ds^2_{\text{AdS}_2} + q_{\mathrm{F}1}^{-1/2}q_{\text{D}3}^{1/2} H_{\text{D}5} H_{\mathrm{NS}5}^{-1/2} dz^2 +H_{\mathrm{NS}5}^{1/2} \,q_{\mathrm{F}1}^{1/2}q_{\text{D}3}^{1/2}q_{\text{D}1}^{-1} d\psi^2+\\
&+ q_{\mathrm{F}1}^{1/2}q_{\text{D}3}^{-1/2}H_{\text{D}5}H_{\mathrm{NS}5}^{1/2} \left(dr^2 + r^2 ds^2_{S^2}\right)+4q_{\mathrm{F}1}^{-1/2}q_{\text{D}3}^{-1/2}H_{\mathrm{NS}5}^{1/2} (dR^2 + \tilde H R^2 ds^2_{\tilde{S}^2}) \,, \\
e^{\Phi} &=  8H_{\text{D}5}^{1/2} H_{\mathrm{NS}5}^{1/4} \,q_{\text{D}1}^{-1/2}q_{\mathrm{F}1}^{-1/4}q_{\text{D}3}^{-3/4} \tilde H^{1/2}\,,\\
H_{(3)} &= -2^{-1}\,q_{\mathrm{F}1}^{1/2}q_{\text{D}3}^{1/2}q_{\text{D}1}^{-1/2}\text{vol}_{\text{AdS}_2} \wedge d\psi+\\
&
+(q_{\mathrm{F}1}q_{\text{D}3}^{-1} H_{\mathrm{NS}5}\,\partial_zH_{\text{D}5}dr-\partial_r H_{\text{D}5}\,dz)\,r^2d\psi \wedge \text{vol}_{S^2}+d((\tilde H-1)R\text{vol}_{\tilde{S}^2})\,,\\
F_{(4)} &= -d[2^{-1}q_{\mathrm{D}1}^{1/2}q_{\mathrm{F}1}^{-1/2}q_{\text{D}3}^{1/2}(3/2+(\tilde H-1)^{-1})R^2 \text{vol}_{\text{AdS}_2} \wedge dz]+\\
&+r^2(4^{-1}q_{\text{F}1}H_{\text{D}5} H_{\mathrm{NS}5} \,dr +R\partial_rH_{\text{D}5}\, dR) \wedge d\psi  \wedge \text{vol}_{ S^2}\\
&-4^{-1} q_{\text{D}3} R^{-1}(\tilde H-1) \, dz \wedge d\psi \wedge \text{vol}_{\tilde S^2}
\,,
\end{aligned}
\end{equation}
where we have defined,
\begin{equation}
\tilde H = \frac{q_{\mathrm{F}1}q_{\text{D}3}}{q_{\mathrm{F}1}q_{\text{D}3} + 16 R^2 H_{\text{NS}5}}\,.
\end{equation}
The fluxes of these solutions are compatible with the brane configurations shown in Table \ref{Table:branesIIA_NATD}. We point out that, as usual for non-Abelian T-dual solutions, a clear prescription to construct the full brane solutions describing the set-ups of Table \ref{Table:branesIIA_NATD} and reproducing \eqref{brane_metric_NATD_0_nh} and \eqref{brane_metric_NATD_pi/2_nh} in the near-horizon limit, is not available.
\begin{table}[http!]
	\renewcommand{\arraystretch}{1}
	\begin{center}
		\scalebox{0.82}[0.8]{
			\begin{tabular}{c||c|cc c c || c  c| c c c}
				branes & $t$ & $\rho$ & $R$ & $\chi^1$ & $\chi^2$ & $z$ & $\psi$ & $r$ & $\theta^1$ & $\theta^2$  \\
				\hline \hline
				$\mrm{D}4$ & $\times$ & $\times$ & $-$ & $\times$ & $\times$ & $-$ & $\times$& $-$ & $-$ & $-$  \\
				$\mrm{D}2$ & $\times$ & $\times$ & $-$ & $-$ & $-$ & $-$ & $\times$& $-$ & $-$ & $-$  \\
				$\mathrm{NS}5$ & $\times$ & $\times$ & $\times$  & $\times$ & $\times$ & $\times$ &$-$& $-$ & $-$ &$-$ \\
				$\mrm{D}2'$ & $\times$ & $-$ & $\times$ & $-$ & $-$ & $ - $ &$\times$& $-$ & $-$ & $-$  \\
				$\mrm{D}4'$ & $\times$ & $-$ & $\times$ & $\times$ & $\times$ & $ - $ &$\times$& $-$ & $-$ & $-$  \\
				$\mrm{F}1$ & $\times$ & $-$ & $-$ & $-$ & $-$ & $\times$ &$-$& $-$ & $-$ & $-$  \\
				$\mrm{D}4''$ & $\times$ & $-$ & $\times$ & $-$ & $-$ & $-$ &$-$& $\times$ & $\times$ & $\times$  \\
				$\mrm{D}6$ & $\times$ & $-$ & $\times$ & $\times$ & $\times$ & $-$ &$-$& $\times$ & $\times$ & $\times$  \\
				$\mathrm{NS}5'$ & $\times$ & $-$ & $-$  & $-$ & $-$ & $\times$ &$\times$& $\times$ & $\times$ &$\times$
			\end{tabular}
			\begin{tabular}{c||c|cc c c || c  c| c c c}
				branes & $t$ & $\rho$ & $R$ & $\chi^1$ & $\chi^2$ & $z$ & $\psi$ & $r$ & $\theta^1$ & $\theta^2$  \\
				\hline \hline
				$\mrm{D}4$ & $\times$ & $\times$ & $-$ & $\times$ & $\times$ & $\times$& $-$ & $-$ & $-$ & $-$  \\
				$\mrm{D}2$ & $\times$ & $\times$ & $-$ & $-$ & $-$ & $\times$ & $-$ & $-$ & $-$ & $-$  \\
				$\mathrm{NS}5$ & $\times$ & $\times$ & $\times$  & $\times$ & $\times$ & $-$& $\times$ & $-$ & $-$ &$-$ \\
				$\mrm{D}2'$ & $\times$ & $-$ & $\times$ & $-$ & $-$ & $\times$&  $ - $ & $-$ & $-$ & $-$  \\
				$\mrm{D}4'$ & $\times$ & $-$ & $\times$ & $\times$ & $\times$ & $\times$& $ - $ & $-$ & $-$ & $-$  \\
				$\mrm{F}1$ & $\times$ & $-$ & $-$ & $-$ & $-$ & $-$& $\times$ & $-$ & $-$ & $-$  \\
				$\mrm{D}4''$ & $\times$ & $-$ & $\times$ & $-$ & $-$ & $-$ &$-$& $\times$ & $\times$ & $\times$  \\
				$\mrm{D}6$ & $\times$ & $-$ & $\times$ & $\times$ & $\times$ & $-$ &$-$& $\times$ & $\times$ & $\times$  \\
				$\mathrm{NS}5'$ & $\times$ & $-$ & $-$  & $-$ & $-$ & $\times$ &$\times$& $\times$ & $\times$ &$\times$
			\end{tabular}
		}
	\end{center}
	\caption{Brane set-ups compatible with the fluxes of the non-Abelian T-dual solutions \eqref{brane_metric_NATD_0_nh} and \eqref{brane_metric_NATD_pi/2_nh}. The coordinates $(R, \chi^1, \chi^2)$ parametrise the open subset of $\mathbb{R}^3$ generated by the action of non-Abelian T-duality on the $S^3$ factor of the Type IIB backgrounds.} \label{Table:branesIIA_NATD}
\end{table}
Nevertheless we can consider their M-theory uplifts. In section \ref{dualitywebs} we discussed the M-theory interpretation of the Abelian T-duals of the Type IIB backgrounds with $\xi=0$ and $\xi=\frac{\pi}{2}$, observing that the two corresponding 11d solutions arise from the same intersection in M-theory (with different smearing of brane charges). Even if for the non-Abelian T-dual backgrounds we do not have full control over the brane solutions behind the AdS$_2$ backgrounds \eqref{brane_metric_NATD_0_nh} and \eqref{brane_metric_NATD_pi/2_nh}, it is possible to show that their M-theory uplifts are related to each other, provided that one makes some assumptions on the spacetime dependence of the function $H_{\text{D}5}$, which implies a particular choice of the charge distribution of branes underlying the non-Abelian T-dual solutions.

The backgrounds \eqref{brane_metric_NATD_0_nh} and \eqref{brane_metric_NATD_pi/2_nh} can be uplifted to M-theory by choosing the same gauge potential for the $F_{(2)}$ flux, namely
\begin{equation}\label{C1NATD}
 C_{(1)}=\frac{q_{\text{D}3}}{8}\,\left(\psi dz-zd\psi\right)\,,
\end{equation}
which is invariant under the following relabeling of the coordinates, $(z, \psi)\rightarrow (\psi, -z)$. The explicit 11d solutions are given in appendix \ref{NATD11d} in equations \eqref{11d_xi=0} and \eqref{11d_xi=pi_2}. We observe that the parameter $q_{\text{D}3}$, whose inclusion in the non-Abelian T-dual backgrounds was discussed in footnote \ref{noteq3}, gains a natural interpretation in M-theory as KK monopole charge.

It was shown in \cite{Lozano:2016kum} that the Abelian T-dual of a certain background can be obtained from the corresponding non-Abelian T-dual one by sending the radial direction of the dual space $\mathbb{R}^3$ to infinity and further compactifying it to the interval $[0,\pi]$. Taking this limit in the
solution \eqref{brane_metric_NATD_0_nh}, we recover the Abelian T-dual of the $\xi=0$ solution \eqref{brane_metric_D3D1F1D5NS5D7_nh_0}, where now $R\in [0,\pi]$. Then we can take the uplift to 11d along the $\chi$ direction, rotate the coordinates as $(\chi, R)\rightarrow (R, -\chi)$ and go back to Type IIA. Doing this we recover the Abelian T-dual of the $\xi=\frac{\pi}{2}$ solution. Such a procedure confirms the reliability of the non-Abelian T-dual backgrounds, since the corresponding Abelian T-duals are shown to be related to the S-dual solutions in Type IIB with $\xi=0$ and $\xi=\frac{\pi}{2}$. Furthermore, the two circular coordinates $(\chi, R)$ parametrise the 2-torus in M-theory that provides the geometrisation of the S-duality transformation in Type IIB.

As it was expected, the 11d uplifts of the solutions \eqref{brane_metric_NATD_0_nh} and \eqref{brane_metric_NATD_pi/2_nh}, given by equations \eqref{11d_xi=0} and \eqref{11d_xi=pi_2} in the Appendix, are not related anymore by a simple rotation of the coordinates as for their Abelian limits. This is reflecting an ``exotic" charge distribution as underlying the intersections depicted in Table \ref{Table:branesIIA_NATD}, which modifies the standard chain of dualities connecting Type IIB string theory to M-theory. Such an ``exotic" charge distribution could be related to the presence of dyonic membranes, which, as shown in \cite{Dibitetto:2020bsh}, define an additional warping between the AdS factor and the internal space, as in \eqref{brane_metric_NATD_0_nh} and \eqref{brane_metric_NATD_pi/2_nh}.


\section{Conclusions}\label{conclusions}

In this paper we have constructed and studied various examples of $\ma N=4$ AdS$_2\times S^3\times S^2\times S^1$ backgrounds fibered over two intervals in Type IIB supergravity. Such solutions have been obtained by extracting the near-horizon limit of a brane solution describing the intersection of D1-F1-D3 branes ending on the D5-NS5-D7 bound state.

As a first example we considered the particular solution for the D5-NS5-D7 bound state reproducing in its near horizon limit the AdS$_6\times S^2\times \Sigma_2$ vacuum with $\Sigma_2$ an annulus. This vacuum geometry is the Abelian T-dual of the Brandhuber-Oz solution of massive Type IIA supergravity. The intersection of D1-F1-D3 branes with the D5-NS5-D7 backreacted geometry gave rise to a non-compact AdS$_2\times S^3$ geometry fibered over a line admitting an asymptotic local description in terms of the AdS$_6\times S^2\times \Sigma_2$ solution. Such a behaviour allowed us to propose an interpretation of the AdS$_2$ solution  as holographically dual to a $\ma N=4$ superconformal quantum mechanics realising a defect within the $\ma N=2$ Sp$(\text{N})$ 5d SCFT dual to the AdS$_6$ geometry.

Secondly, we focused on the particular subclass of $\ma N=4$ AdS$_2\times S^3\times S^2\times S^1$ solutions fibered over two intervals featured by the absence of D7 branes. Even if this requirement implies that the defect interpretation in AdS$_6$ is lost, this subclass is interesting since we could act locally with an SL$(2, \mathbb{R})$ transformation to generate a vast class of inequivalent backgrounds parametrised by a continuous parameter $\xi \in [0,\frac{\pi}{2}]$. 

We then focused on the two S-dual backgrounds with $\xi=0$ and $\xi=\frac{\pi}{2}$, and studied their Type IIA realisation by acting with Abelian T-duality along the $S^1$ present in both backgrounds. In this way we constructed the entire chain of dualities providing the M-theory origin of our S-dual pair of solutions. This allowed us to show that they belong to the general class of $\ma N=(0,4)$ AdS$_3$ solutions to M-theory classified in \cite{Lozano:2020bxo}. Remarkably, we showed that the T-dual of the $\xi=\frac{\pi}{2}$ solutions is related to the AdS$_3\times S^3 \times S^3$ backgrounds studied in \cite{Faedo:2020nol}, which were shown to asymptote locally to the AdS$_7/\mathbb{Z}_k\times S^4$ vacuum geometry of M-theory. Thus, in the absence of D7-branes we lost the line defect interpretation within AdS$_6$ in Type IIB, but we recovered a surface defect interpretation within the $\ma N=(1,0)$ 6d SCFT dual to the AdS$_7/\mathbb{Z}_k$ solution in M-theory.

We concluded by deriving the non-Abelian T-duals in Type IIA of the S-dual pairs with $\xi=0$ and $\xi=\frac{\pi}{2}$ and discussing their embeddings in M-theory. 

Our results in this paper contribute to deepen our understanding of the interrelations between $\ma N=4$ AdS$_2$ and $\ma N=(0,4)$ AdS$_3$ solutions to Type II and M theories with  4 (small)  supersymmetries. In this scenario there are two research directions that we think would be interesting to pursue in the future.
First, it would be interesting to construct a more general and systematic classification of $\ma N=4$ AdS$_2\times S^3$ solutions to Type IIB supergravity, in particular including an additional warping between the AdS$_2$ and the $S^3$ factors. One could try to search for these solutions in lower dimensional gauged supergravities, as initiated in \cite{Dibitetto:2018gtk}. These more general backgrounds would be described in term of a brane intersection involving dyonic membranes, as it has been highlighted in M-theory for AdS$_3\times S^3$ backgrounds \cite{Dibitetto:2020bsh}.
 A second interesting research direction is the construction of the quiver defining the superconformal quantum mechanics dual to the AdS$_2$ solution studied in subsection \ref{defects}, following the ideas of \cite{Faedo:2020nol,Lozano:2022ouq,Lozano:2022vsv}. Such a field theory would explicitly describe a conformal line defect within the 5d SCFT dual to the AdS$_6\times S^2 \times \Sigma_2$ background emerging in the asymptotics.

\section*{Acknowledgements}

YL and CR are partially supported by the AEI through the Spanish grant MCIU-22-PID2021-123021NBI00 and by the FICYT through the Asturian grant SV-PA-21-AYUD/2021/52177. CR is  supported by a Severo Ochoa Fellowship by the Principality of Asturias (Spain).
The work of NP is supported by the Israel Science Foundation (grant No. 741/20) and by the German Research Foundation through a German-Israeli Project Cooperation (DIP) grant ``Holography and the Swampland".

\appendix

\section{M-theory uplifts of the non-Abelian T-dual backgrounds}\label{NATD11d}

In this Appendix we provide the M-theory uplift of the two backgrounds \eqref{brane_metric_NATD_0_nh} and \eqref{brane_metric_NATD_pi/2_nh} obtained by acting with non-Abelian T-duality on the S-dual solutions \eqref{brane_metric_D3D1F1D5NS5D7_nh_0} and \eqref{brane_metric_D3D1F1D5NS5D7_nh_pi_2}. If one introduces the gauge potential \eqref{C1NATD},
the 11d uplift of the background \eqref{brane_metric_NATD_0_nh} is given by,
\begin{align}
\begin{split}  \label{11d_xi=0}
ds_{11}^2 
&=  16^{-1} q_{\text{D}1}^{2/3}q_{\mathrm{F}1}^{1/3}q_{\text{D}3}H_{\text{D}5}^{-2/3} H_{\mathrm{NS}5}^{-1/3} H^{-1/3} ds^2_{\text{AdS}_2} + 4^{-1} q_{\text{D}1}^{2/3}q_{\text{D}3} q_{\mathrm{F}1}^{-2/3} H_{\text{D}5}^{1/3} H_{\mathrm{NS}5}^{-1/3} H^{-1/3} dz^2 +\\
&+ 4^{-1} q_{\text{D}1}^{-1/3}q_{\mathrm{F}1}^{1/3}q_{\text{D}3} H_{\text{D}5}^{-2/3}H_{\mathrm{NS}5}^{2/3}  H^{-1/3} d\psi^2 +4^{-1} q_{\text{D}1}^{2/3}q_{\mathrm{F}1}^{1/3} H_{\text{D}5}^{1/3}H_{\mathrm{NS}5}^{2/3} H^{-1/3} \left(dr^2 + r^2 ds^2_{S^2}\right)+\\
&+  q_{\text{D}1}^{-1/3}q_{\mathrm{F}1}^{1/3}H_{\text{D}5}^{1/3} H_{\mathrm{NS}5}^{-1/3}  (H^{-1/3} dR^2 + H^{2/3} R^2 ds^2_{\tilde{S}^2}) 
+\\
&+
 16 q_{\text{D}1}^{-1/3}q_{\mathrm{F}1}^{-2/3}q_{\text{D}3}^{-1} H_{\text{D}5}^{1/3} H_{\mathrm{NS}5}^{2/3} H^{2/3}  (d\chi + 8^{-1}q_{\text{D}3}(\psi\,dz-z\,d\psi))^2\,,
 \\
G_{(4)} &= d[2^{-1}q_{\text{D}1}^{-1/2}q_{\mathrm{F}1}^{1/2}q_{\mathrm{D}3}^{1/2}(3/2+(H-1)^{-1})R^2 \text{vol}_{\text{AdS}_2} \wedge d\psi]\\
&+4 q_{\text{D}1}^{-1} HR^3H_{\text{D}5} \, dz \wedge d\psi \wedge \text{vol}_{\tilde S^2}-4^{-1}q_{\text{D}1}H_{\text{D}5} H_{\mathrm{NS}5}\,r^2 \,dz\wedge dr \wedge \text{vol}_{ S^2}+\\
&+r^2R(q_{\mathrm{F}1} q_{\mathrm{D}3}^{-1}H_{\mathrm{NS}5}\,\partial_zH_{\text{D}5}dr-\partial_r H_{\text{D}5}\,dz)\,  \wedge\text{vol}_{ S^2}\wedge dR + [-2^{-1}\,q_{\text{D}1}^{1/2} q_{\mathrm{D}3}^{1/2} q_{\mathrm{F}1}^{-1/2} \text{vol}_{\text{AdS}_2} \wedge dz+\\
&+\,\partial_r H_{\mathrm{NS}5}\,r^2d\psi \wedge \text{vol}_{S^2}+d((H-1)R\,\text{vol}_{\tilde{S}^2})] \wedge (d\chi + 8^{-1}q_{\text{D}3}(\psi\,dz-z\,d\psi)).
\end{split}
\end{align}
Using the same gauge potential \eqref{C1NATD} and uplifting the \eqref{brane_metric_NATD_pi/2_nh} solution, we obtain
\begin{align} \label{11d_xi=pi_2}
\begin{split} 
ds_{11}^2 &= 16^{-1} q_{\text{D}1}^{1/3} q_{\mathrm{F}1}^{2/3}q_{\text{D}3} H_{\text{D}5}^{-1/3} H_{\mathrm{NS}5}^{-2/3} \tilde H^{-1/3} ds^2_{\text{AdS}_2} + 4^{-1} q_{\text{D}1}^{1/3} q_{\mathrm{F}1}^{-1/3}q_{\text{D}3} H_{\text{D}5}^{2/3} H_{\mathrm{NS}5}^{-2/3}\tilde  H^{-1/3} dz^2 +\\
&+ 4^{-1} q_{\text{D}1}^{-2/3} q_{\mathrm{F}1}^{2/3}q_{\text{D}3} H_{\text{D}5}^{-1/3} H_{\mathrm{NS}5}^{1/3}\tilde H^{-1/3} d\psi^2 + 4^{-1} q_{\text{D}1}^{1/3} q_{\mathrm{F}1}^{2/3} H_{\text{D}5}^{2/3}H_{\mathrm{NS}5}^{1/3} \tilde H^{-1/3} \left(dr^2 + r^2 ds^2_{S^2}\right)+\\
&+ q_{\text{D}1}^{1/3} q_{\mathrm{F}1}^{-1/3} H_{\text{D}5}^{-1/3} H_{\mathrm{NS}5}^{1/3} (\tilde H^{-1/3} dR^2 + \tilde H^{2/3}  R^2 ds^2_{\tilde{S}^2}) +\\
&+ 16  q_{\text{D}1}^{-2/3}q_{\mathrm{F}1}^{-1/3}q_{\text{D}3}^{-1}  H_{\text{D}5}^{2/3} H_{\mathrm{NS}5}^{1/3}\tilde H^{2/3} (d\chi + 8^{-1}q_{\text{D}3}(\psi\,dz-z\,d\psi))^2\,,
\\
G_{(4)} &= -d[2^{-1}q_{\mathrm{D}1}^{1/2}q_{\mathrm{F}1}^{-1/2}q_{\text{D}3}^{1/2}(3/2+(\tilde H-1)^{-1})R^2 \text{vol}_{\text{AdS}_2} \wedge dz]+r^2(4^{-1}q_{\text{F}1}H_{\text{D}5} H_{\mathrm{NS}5} \,dr +\\
&+R\partial_rH_{\text{D}5}\, dR) \wedge d\psi  \wedge \text{vol}_{ S^2}-4^{-1} q_{\text{D}3} R^{-1}(\tilde H-1) \, dz \wedge d\psi \wedge \text{vol}_{\tilde S^2}\\
&+[-2^{-1}\,q_{\mathrm{F}1}^{1/2}q_{\text{D}3}^{1/2}q_{\text{D}1}^{-1/2}\text{vol}_{\text{AdS}_2} \wedge d\psi
+
(q_{\mathrm{F}1}q_{\text{D}3}^{-1} H_{\mathrm{NS}5}\,\partial_zH_{\text{D}5}dr-\partial_r H_{\text{D}5}\,dz)\,r^2d\psi \wedge \text{vol}_{S^2}\\
&+d((\tilde H-1)R\text{vol}_{\tilde{S}^2})] \wedge (d\chi + 8^{-1}q_{\text{D}3}(\psi\,dz-z\,d\psi)).
\end{split}
\end{align}

\end{document}